\documentclass[twocolumn,prb,aps,floatfix,superscriptaddress,nobibnotes]{revtex4-2}
\usepackage{graphicx}
\usepackage{enumerate}
\usepackage[colorlinks=true,citecolor=blue]{hyperref}
\usepackage{textcomp}
\usepackage{amsmath}
\usepackage{amssymb}
\usepackage{pgf}
\usepackage{svg}
\usepackage{subfigure}
\usepackage[english]{babel}
\usepackage[normalem]{ulem}
\usepackage{tabularray}
\usepackage{xcolor}
\usepackage{array}
\usepackage{comment}

\usepackage{dcolumn}
\usepackage{bm}
\usepackage{amsfonts}
\usepackage{siunitx}
\usepackage{dsfont}
\usepackage{placeins}

\DeclareUnicodeCharacter{03BD}{$\nu$}

\definecolor{bluepoli}{RGB}{0,36,179}
\definecolor{redpoli}{RGB}{204,0,51}
\definecolor{greenpoli}{RGB}{45,137,0}
\definecolor{purplepoli}{RGB}{153,102,204}
\definecolor{azzurropoli}{RGB}{51,53,204}
\definecolor{orangepoli}{RGB}{255,124,17}
\usepackage{hyperref}
\hypersetup{
	colorlinks,
	linkcolor={bluepoli}, 
	citecolor={redpoli}, 
	urlcolor={redpoli} %
}

\usepackage[capitalise]{cleveref}

\begin{document}

\preprint{APS/123-QED}

\title{Automated \textit{in situ} optimization and disorder mitigation in a quantum device}

\author{Jacob Benestad}
\thanks{These authors contributed equally}
\affiliation{Center for Quantum Spintronics, Department of Physics, Norwegian University of Science and Technology, NO-7491 Trondheim, Norway}
\author{Torbjørn Rasmussen}%
\thanks{These authors contributed equally}
\affiliation{Center for Quantum Devices, Niels Bohr Institute, University of Copenhagen, 2100 Copenhagen, Denmark}
\affiliation{QuTech and Kavli Institute of Nanoscience, Delft University of Technology, 2600 GA Delft, The Netherlands}

\author{Bertram Brovang}
\thanks{These authors contributed equally}
\affiliation{Center for Quantum Devices, Niels Bohr Institute, University of Copenhagen, 2100 Copenhagen, Denmark}

\author{Oswin Krause}
\affiliation{Department of Computer Science, University of Copenhagen, 2100 Copenhagen, Denmark}

\author{Saeed~Fallahi}
\affiliation{Department of Physics and Astronomy, Purdue University, West Lafayette, Indiana 47907, USA}
\affiliation{Birck Nanotechnology Center, Purdue University, West Lafayette, Indiana 47907, USA}

\author{Geoffrey~C.~Gardner}
\affiliation{Birck Nanotechnology Center, Purdue University, West Lafayette, Indiana 47907, USA}

\author{Michael~J.~Manfra}
\affiliation{Department of Physics and Astronomy, Purdue University, West Lafayette, Indiana 47907, USA}
\affiliation{Birck Nanotechnology Center, Purdue University, West Lafayette, Indiana 47907, USA}
\affiliation{Elmore Family School of Electrical and Computer Engineering, Purdue University, West Lafayette, Indiana 47907, USA}
\affiliation{School of Materials Engineering, Purdue University, West Lafayette, Indiana 47907, USA}
	
\author{Charles M. Marcus}
\affiliation{Center for Quantum Devices, Niels Bohr Institute, University of Copenhagen, 2100 Copenhagen, Denmark}

\author{Jeroen~Danon}
\affiliation{Center for Quantum Spintronics, Department of Physics, Norwegian University of Science and Technology, NO-7491 Trondheim, Norway}

\author{Ferdinand Kuemmeth}
\affiliation{Center for Quantum Devices, Niels Bohr Institute, University of Copenhagen, 2100 Copenhagen, Denmark}
\affiliation{Institute of Experimental and Applied Physics, University of Regensburg, 93040 Regensburg, Germany}

\author{Anasua Chatterjee}
\email{Anasua.Chatterjee@tudelft.nl}
\affiliation{Center for Quantum Devices, Niels Bohr Institute, University of Copenhagen, 2100 Copenhagen, Denmark}
\affiliation{QuTech and Kavli Institute of Nanoscience, Delft University of Technology, 2600 GA Delft, The Netherlands}

\author{Evert van Nieuwenburg}
\email{e.p.l.van.nieuwenburg@liacs.leidenuniv.nl}
\affiliation{$\langle aQa^{L} \rangle$ at Lorentz Institute and Leiden Institute of Advanced Computer Science, Leiden University, P.O. Box 9506, 2300 RA Leiden, The Netherlands}

\date{\today}

\begin{abstract}

We investigate automated \textit{in situ} optimization of the potential landscape in a quantum point contact device, using a $3 \times 3$ gate array patterned atop the constriction.
Optimization is performed using the covariance matrix adaptation evolutionary strategy, for which we introduce a metric for how ``step-like'' the conductance is as the channel becomes constricted.
We first perform the optimization of the gate voltages in a tight-binding simulation and show how such \textit{in situ} tuning can be used to mitigate a random disorder potential.
The optimization is then performed in a physical device in experiment, where we also observe a marked improvement in the quantization of the conductance resulting from the optimization procedure.
\end{abstract}

\maketitle


\paragraph*{Introduction}

Machine learning has garnered increasing attention within quantum and condensed matter physics in recent years, finding applications as a new and useful tool for scientific discovery~\cite{neupert_introduction_2022, dawid_modern_2023, gebhart_learning_2023}. One such application is the facilitation of quantum experiments by partly or fully automating the experimental workflow in cases where tuning of several parameters is necessary~\cite{ares_machine_2021,Zwolak2023}. Significant work has already been done towards characterization of multi-quantum-dot devices~\cite{nguyen_deep_2021, Zubchenko2024, Kalantre2019, krause_learning_2022, krause_estimation_2022, koch_adversarial_2023, taylor_neural_2024, van_driel_cross-platform_2024} and automated tuning of such systems to a desired operating configuration~\cite{moon_machine_2020, durrer_automated_2020, Ziegler2023}. One common challenge in tuning up quantum devices is posed by material imperfections and other types of unpredictable disorder, which often hinder straightforward navigation through a large parameter space.
An important envisioned application of machine learning is thus to characterize or mitigate the disorder in quantum devices, yielding optimized device tunings that would be difficult to obtain manually~\cite{thamm_machine_2023, craig_bridging_2024, chatzikyriakou_unveiling_2022, benestad_2024}.

In this work, we implement an automated evolutionary strategy for \textit{in situ} optimization of a quantum device.
We demonstrate how our tuning protocol can both enhance device functionality and mitigate the effects of disorder.
As an illustrative proof-of-principle, we apply the same optimization algorithm to a numerically simulated device as well as to a real device in experiment, where in both cases we choose the simplest approach resulting in efficient optimization.

We focus on a quantum device for which most of the physics is well understood, namely a quantum point contact (QPC) defined within a GaAs-based two-dimensional electron gas, where the potential landscape in the constriction can be modified via a $3 \times 3$ array of electrostatic top-gates.
QPCs have proved to be an invaluable resource for development of semiconductor quantum technologies.
The quantized conductance demonstrated in such devices \cite{wharam_one-dimensional_1988, van_wees_quantized_1988, buttiker_quantized_1990} is not only a fundamental hallmark of quantum physics, but has also found important applications in quantum sensing: the sensitivity to nearby charges when operated between two conductance plateaus makes QPCs well suited for detection of single-electron tunneling events in quantum-dot systems \cite{field_measurements_1993, elzerman_single-shot_2004, petta_coherent_2005, reilly_fast_2007, cassidy_single_2007, gustavsson_electron_2009, barthel_rapid_2009, bauerle_coherent_2018}, which is an important ingredient in readout of spin qubits \cite{vandersypen_interfacing_2017, burkard_semiconductor_2023}.
Other uses for QPCs are for instance to build interferometers for probing anyonic statistics \cite{bishara_interferometric_2009, rosenow_current_2016, bartolomei_fractional_2020, nakamura_fabry-perot_2023}, or to make switches to toggle between tunneling and ballistic transport~\cite{smith_quantum_1995}.

To optimize QPC functionality, i.e., the appearance of stable plateaus with integer quantized conductance, separated by sharp steps, we use the ``covariance matrix adaptation evolution strategy'' (CMA-ES)~\cite{hansen_cma_2023} algorithm. 
Since this algorithm is (i) gradient-free, (ii) extremely simple to implement, and (iii) relatively insensitive to device-specific details influencing the loss landscape for optimization, it combines several desirable characteristics in the context of device tuning where measurements include noise and the optimization landscapes are likely non-trivial.
Via the nine gate voltages of the $3 \times 3$ ``pixel''-gate array we have direct control over the potential landscape, and can hence correct for possible microscopic disorder at length scales comparable to the pixel-gate dimensions. 
We first implement the optimization routine on a simulated device (using \texttt{KWANT}~\cite{groth_kwant_2014}), where we have full control over the properties of the device and disorder, and can experiment with possible functionality metrics.
The resulting optimized gate configuration, both with and without random disorder, yields clear quantized conductance plateaus at values consistent with the physics of QPC devices, even as the optimization algorithm remains agnostic about the underlying disorder distribution, and about where the conductance plateaus should occur.
More importantly, we then employ the same numerical optimization routine directly on a real device, and also demonstrate \textit{in-situ} mitigation of device-specific disorder in an experiment. 



\FloatBarrier

\paragraph*{Device to optimize}
\label{Device_to_optimize}
It is well known that ballistic electron transport through nanowires or 2DEG split-gate structures occurs in quantized conductance steps of $2e^2/h$.
For a sufficient splitting of the energy subbands, this becomes visible in experiments as conductance plateaus with sharp jumps between them when the transverse confinement is changed \cite{van_wees_quantized_1988, wharam_one-dimensional_1988}.
Except in the ideal case at zero temperature, these tell-tale steps in the quantized conductance curve will to some extent be smoothed out and distorted due to factors such as temperature, scattering off the confining potential, and disorder \cite{thomas_ihn_semiconductor_2010}. A question one might thus ask is if it is possible to construct a tunable potential landscape to improve the staircase signature of the quantized conductance.

\begin{figure}
    \centering
    \includegraphics[width=\linewidth]{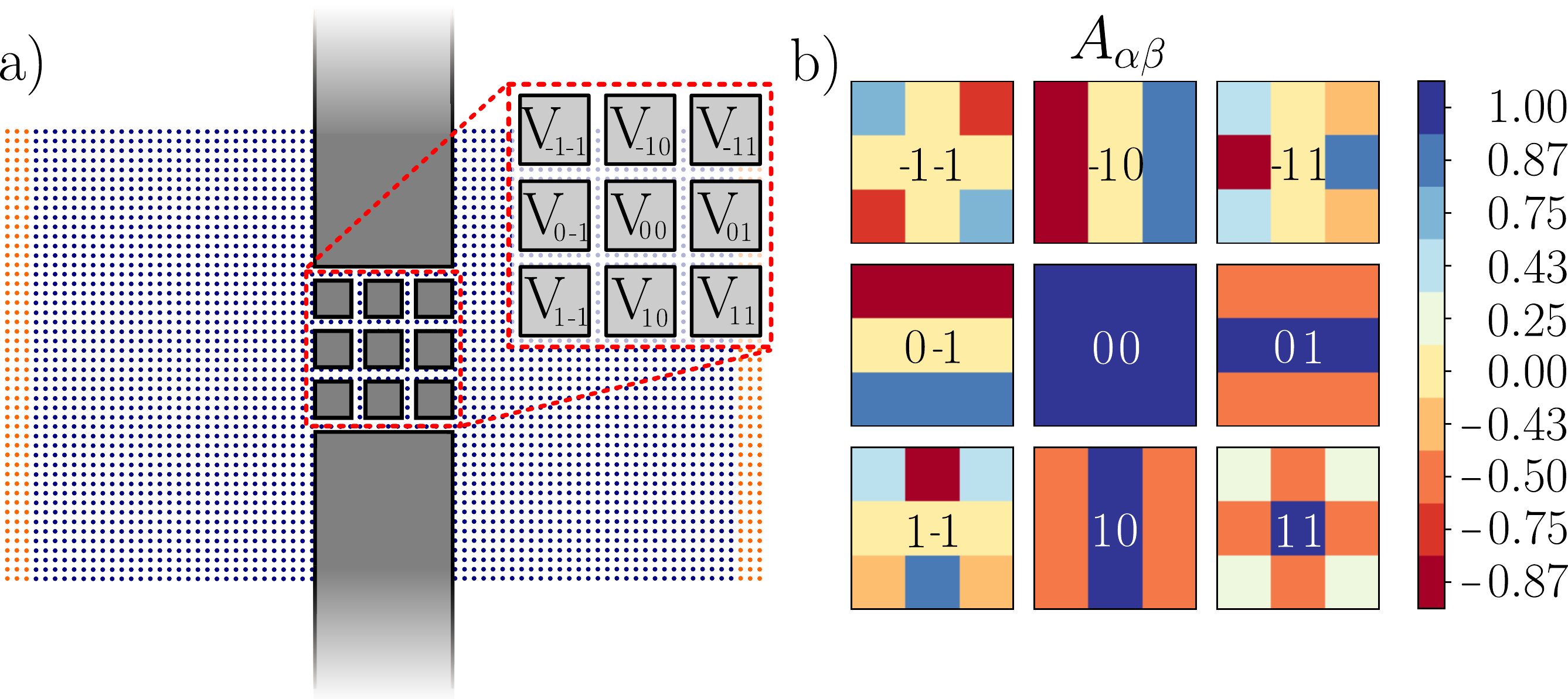}
    \caption{(a) Cartoon of the gate layout over a tight-binding scattering region. (b) Fourier modes $A_{\alpha\beta}$ of the pixel gate voltages.}
    \label{fig:qpc_layout}
\end{figure}

We consider the following system for optimization of the quantized conductance curve: a $3 \times 3$ array of square voltage-controlled gates, analogous to pixels in a camera sensor, patterned between a split-gate structure of two outer gates. 
Together, these $11$ gates lie some distance $z$ above a 2DEG in an AlGaAs/GaAS heterostructure, and the electric potential from the applied voltages shapes the potential landscape in the 2DEG. 
A sketch of the gate pattern is shown in \cref{fig:qpc_layout}(a). 
We denote the pixel gate voltages as $V_{\eta \nu}$ (with $\eta, \nu \in \{-1,0,1\}$ denoting the pixel position), while the split-gate voltages are named $V_{\text{SG}1}$ and $V_{\text{SG}2}$.

Having several tunable gates results in many choices for what gate voltages to sweep over to constrict the channel to produce a conductance staircase. 
We will denote the parameter for constricting the QPC transport channel as $V_{\text{QPC}}$, which is the independent variable of the conductance curve. 
Different choices will result in different behavior; the results of sweeping the outer gates for instance may be of limited quality due to the large distance between them. 
Other options include keeping the outer gates fixed and sweeping over a pair of pixel-gate voltages in the upper and lower row, such as $V_{-10}$ and $V_{10}$, see \cref{fig:qpc_layout}(a), or sweeping over the average gate voltage of all the pixels.
The choice of $V_{\text{QPC}}$ has a large impact on the quantized conductance features, since it sets both the effective length and width of the QPC~\cite{Iqbal_QPC_2013}. 
For now, we will keep the outer gates $V_{\text{SG}1}$ and $V_{\text{SG}2}$ at fixed values, as they are too far apart to form a QPC on their own and we will not specify $V_{\text{QPC}}$ yet. 
The optimization is done on the pixel gates $V_{\eta \nu}$, either directly on the pixels not used for $V_{\text{QPC}}$, or through the Fourier modes presented in the next section.

\begin{figure}
    \centering
    \includegraphics[width=\linewidth]{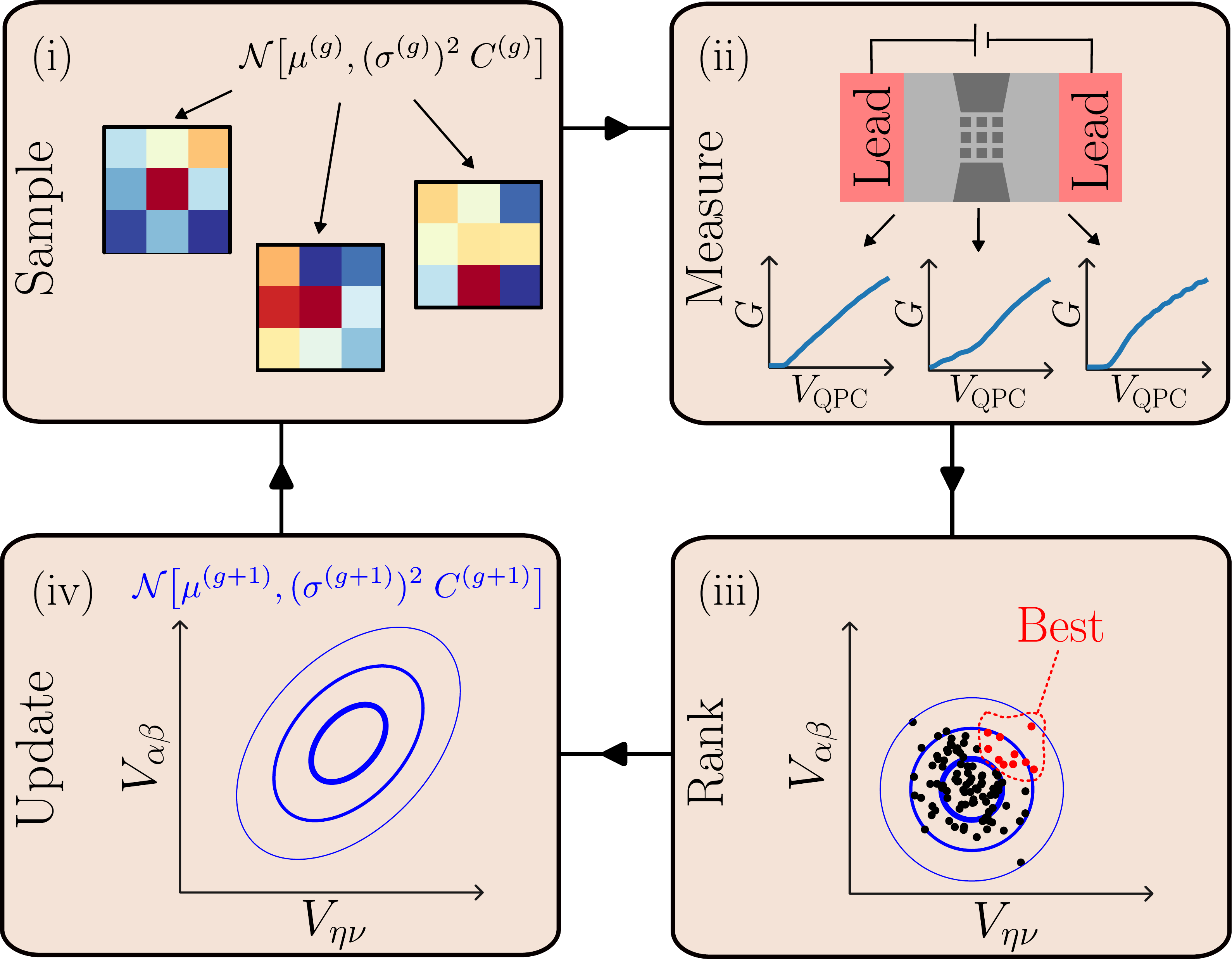}
    \caption{Schematic of the CMA-ES algorithm interfaced to either a simulation or an actual QPC experiment to optimize the pixel-gate voltages. (i) Configurations are sampled, (ii) tested in experiment/simulation, (iii) ranked according to \cref{loss} and (iv) the population is updated.}
    \label{fig:cmaes}
\end{figure}

For the optimization of the conductance curve, we use the CMA-ES algorithm~\cite{hansen_cma_2023} (via the \texttt{CMA} python library), where a population of samples drawn from a multivariate normal distribution (MvN) is ranked according to their ``fitness'', with a fraction of the fittest samples becoming the basis for the next population. Specifically, at generation $g$ of the algorithm, a number of $n_{\text{pop}}$ candidates for the set of tunable parameters $\bm{V}^{(g+1)} \sim \mathcal{N}\left(\bm{\mu}^{(g)}, \sigma^{2\,(g)} C^{(g)} \right)$ are drawn. The drawn samples are all evaluated by a loss function $L\left(\bm{V}^{(g+1)}\right)$, where $n_{\text{fit}}$ ( $1 \leq n_{\text{fit}} < n_{\text{pop}}$) samples giving the smallest loss are used to update the MvN distribution for the population. The mean $\bm{\mu}^{(g+1)}$ of the new MvN is a weighted average of the fittest samples, the ``size'' of the search area $\sigma^{2\,(g+1)}$ is increased or decreased depending on how far away from the old MvN mean value the fittest samples are located, and the covariance matrix $C^{(g+1)}$ of the new MvN is updated according to what direction the fittest samples are located in comparison to the population as a whole. A schematic of how CMA-ES can be implemented for optimization of the QPC array is shown in \cref{fig:cmaes}. Being a stochastic and derivative-free method, CMA-ES is a well-suited method for the task at hand since the relationship between the gate voltages and the shape of the conductance curve cannot be written analytically in general, especially when considering the randomness associated with disorder.

To quantify how ``step-like'' the conductance $G[k]$ looks when sweeping over values of $V_{\text{QPC}}[k]$ (with $k = 1,2,..., M$ data points), we define a loss function $L$ that favors long regions of small derivatives while penalizing a steadily growing conductance. 
This is achieved by summing up the cube root of the distance between adjacent conductance values
\begin{equation}
\label{loss}
    L = \frac{\sum_{k=1}^{M-1}\sqrt[3]{|G[k+1]-G[k]|+\varepsilon}}{1+N_{G>\text{thr}}/M},
\end{equation}
where the cube root yields a preference of a few large terms over many small terms, due to the reverse Minkowski inequality.
A small offset value of $\varepsilon=0.02\,(e^2/h)$ is added to reduce the effect of noise, and the term $N_{G>\text{thr}}/M$ in the denominator penalizes closing of the QPC completely. Here, $N_{G>\text{thr}}$ counts the number of conductance values exceeding a threshold [$2 \times 10^{-5}\, (e^2/h)$].
We further constrain the allowed range of pixel gate voltages $V_{\eta \nu}$, and any out-of-bounds voltages are projected to the nearest in-bounds value while getting a penalty proportional to the square of the distance from the allowed range.
Importantly, the loss function does not include any information about the quantized conductance plateau values. In other words, solutions are determined by the physics of quantized conductance rather than any type of ``gate-engineered overfitting''.

\FloatBarrier

\paragraph*{Simulations}
\label{Simulations}
We simulate the transport properties of our QPC array device using the software package \texttt{KWANT}, which provides numerical solutions of tight-binding models \cite{groth_kwant_2014}. We model the 2DEG with gate-defined potentials $E_i(x,y)$ with $i\in\{\text{SG1},\, \text{SG2},\, (\eta, \nu)\}$ and disorder $E_{\text{dis}}(x,y)$ by the continuous Hamiltonian
\begin{equation}
\label{ham}
    H = -\frac{\hbar^2}{2m^\ast}\big(\partial_x^2+\partial_y^2\big) - \mu + \sum_i E_{i}(x,y) + E_{\text{dis}}(x,y),
\end{equation}
where $m^\ast$ is the effective mass and $\mu$ the electrochemical potential.
The potentials $E_i(x,y)$ are calculated assuming rectangular gates at a given distance above the 2DEG~\cite{davies_modeling_1995, foulk_theory_2024}, ignoring any dielectric screening in the material, see the Supplementary Material for more details.
For the disorder potential $E_{\text{dis}}(x,y)$ we draw
random onsite disorder values $u_{\text{dis}}(x,y) \sim \text{uniform}(0, V_{\text{dis}})$ and set a correlation length $\lambda_{\text{dis}}$ for the disorder by filtering out short-wavelength modes \cite{thamm_machine_2023},
\begin{equation}
    E_{\text{dis}}(x,y) = \mathcal{F}^{-1}\left[ e^{-2\pi |\bm{q}|\lambda_{\text{dis}}} \mathcal{F}\left[ u_{\text{dis}}(x,y) \right]\right],
\end{equation}
where $f({\bf q}) = {\mathcal F}[ f({\bf r}) ]$ is the Fourier transform of $f({\bf r})$.

Through discretization of the Hamiltonian \cref{ham} on a $500\times 300$ lattice of grid spacing $a=\SI{10}{\nano\meter}$ we define the tight-binding hopping parameter $t=\hbar^2/2m^\ast a^2$ to be used in the numerical calculations, with the same value used for hopping to the leads (orange regions in panel (a) of Fig.~\ref{fig:qpc_layout}). 
For the effective mass, we choose $m^\ast = 0.067m_{\text{e}}$, with $m_{\text{e}}$ being the electron mass, we set the electrochemical potential to $\mu=t/4$, and we assume zero temperature. 
The conductance is then found by calculation of the scattering matrix and thereby the transmission eigenvalues.

Instead of optimizing the pixel voltages $V_{\eta \nu}$ directly with the CMA-ES algorithm, we optimize the Fourier modes $A_{\alpha\beta}$ (shown in panel (b) of \cref{fig:qpc_layout}), defined by
\begin{equation}
\begin{split}
V_{\eta\nu} = \sum_{\alpha,\beta}
A_{\alpha\beta} {} & {} \cos \left( \frac{\pi \alpha}{12} [ 8\eta-3(\alpha-1)] \right) \\
{} & {} \quad\times \cos \left( \frac{\pi \beta}{12} [ 8\nu+3(\beta-1)] \right),
\end{split}
\end{equation}
where $\alpha,\beta \in \{-1,0,1\}$,
and we chose the QPC sweep parameter to be the average pixel voltage $A_{00} = V_{\text{QPC}}$.
Tuning voltages collectively via Fourier coefficients is more robust than tuning individual gates since we are tuning overall ``shapes'' of the 2DEG potential rather than individual areas~\cite{thamm_machine_2023}.
Our choice of $A_{00} = V_{\text{QPC}}$ is motivated by $A_{00}$ being the average, so that for our simulations we can sweep $V_{\text{QPC}}$ from $-1.75\,\mu$ to $0$, and set the pixel-gate bounds to be $V_{\eta\nu}/\mu\in(-3,1)$.
Here $n_{\text{pop}}=56$, chosen based on the number of CPU cores.

\begin{figure}
    \centering
    \includegraphics[width=\linewidth]{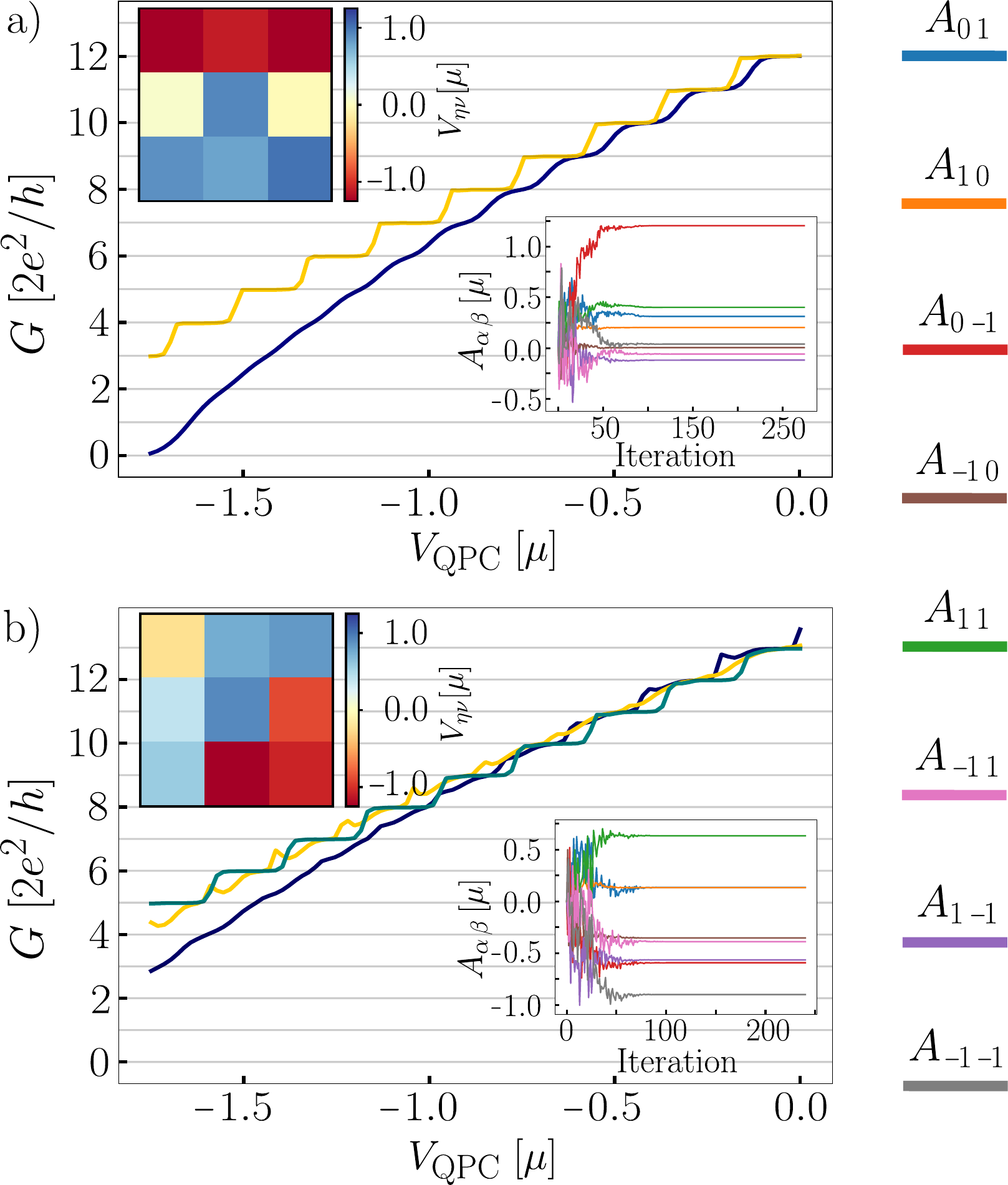}
    \caption{(a) Conductance for a simulated device without disorder as a function of $V_ {\text{QPC}} = A_{0 0}$, for the case with all other Fourier coefficients $A_{\alpha \beta} = 0$ (blue) and after optimization (yellow). (b) Conductance as a function of $V_ {\text{QPC}}$ in the presence of disorder characterized by $V_{\text{dis}}=0.352\,\mu$ and $\lambda_{\text{dis}}=\SI{300}{\nano\meter}$. In addition to traces with the same two Fourier coefficient configurations as in (a) (blue,yellow) we show the optimized trace in the presence of the disorder (turquoise).
    (right insets) Convergence of the Fourier coefficients (color coding shown on the right) during the optimization. (left insets) The final pixel configuration around the average. }
    \label{fig:qpc_learning}
\end{figure}

\paragraph*{Simulation results}
\label{Simulation_results}
Optimization was performed for a simulated QPC device both in the absence and presence of disorder.
In \cref{fig:qpc_learning}(a) we plot the conductance as a function of the average pixel voltage $V_{\text{QPC}} = A_{00}$ for a disorder-free system in the cases of all other Fourier components being either set to zero (blue) or optimized (yellow). 
Both curves have a step-like shape, with sharper features in the optimized case.
The left inset shows the final optimized pixel-gate voltages. 
Interestingly, the optimized solution is asymmetric~\footnote{Additional optimization runs show that the algorithm also finds the mirror image solution, see Supplementary Materials.}.
In \cref{fig:qpc_learning}(b) we investigate the case with disorder.
We set the correlation length to be $\lambda_{\text{dis}}=\SI{300}{\nano\meter}$, comparable to the pixel gate sizes (for other choices of $\lambda_{\text{dis}}$ see discussion in Supplement). The strength of the disorder was set to $V_{\text{dis}}=0.352\,\mu$, a value large enough so that the staircase for the optimal solution without disorder was visibly degraded after adding the disorder. 
We show the conductance for a fixed disorder configuration for three different pixel-gate configurations: the two same configurations as in \cref{fig:qpc_learning}(a) (blue, yellow) and in addition a configuration that was optimized in the presence of disorder (turquoise).
The final optimized configuration is again shown in the left inset.

The right insets in \cref{fig:qpc_learning} show the evolution of the eight variables as a function of iteration number during optimization.
We see that the algorithm initially probes a relatively large search space but eventually converges towards a voltage configuration that gives a clear conductance staircase. In both cases, the optimization loop was terminated after reaching the maximal time limit of $48$ hours, rather than converging within the \texttt{CMA} library's default tolerances on change in loss function and optimized parameters~\cite{hansen_cma_2023}.

Already without disorder, an optimized potential landscape seems to yield conductance steps that are much more pronounced than when only the average pixel values are swept. It appears that the visibility of plateaus in the unoptimized case becomes lower as the channel becomes more constrained, which could be due to differences in how much the potential shape is changed along the $x$ and $y$ directions as the pixel gate voltages are swept. When tuning in the presence of disorder, the staircase that had vanished due to the disorder now reappears after tuning the gates to a new set of values, indicating that it can adapt to a certain extent to the randomness. We note that different spatial configurations of the random disorder affect the disorder-free solution to different degrees; sometimes the impact is small, and sometimes even the completely unoptimized sweep shows a clearer staircase than without disorder. However, when there is disorder, the optimized devices nearly always outperform those configured for no disorder or without any optimization at all (see Supplementary). Lastly, we mention the observation that all the plateaus occur on integer conductance values without having coded this as a constraint into the loss function, suggesting that the algorithm is not just finding voltage configurations that happen to produce plateaus, but that the solution found is indeed a manifestation of QPC physics.

\begin{figure*}
    \centering
    \includegraphics{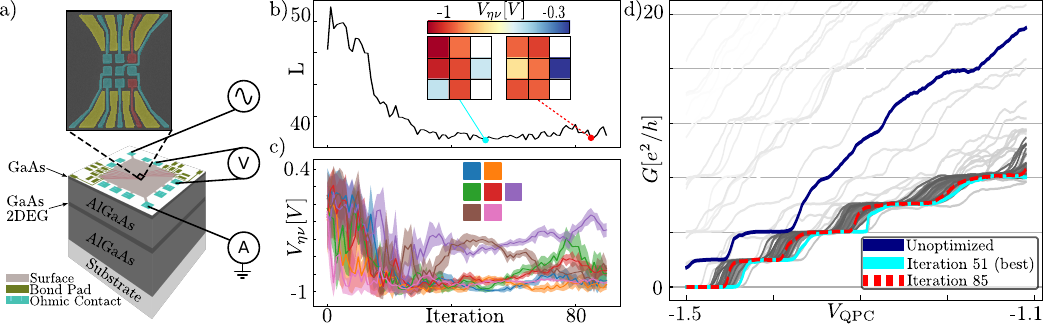}
    \caption{(a) Device heterostructure and false-colored SEM image, indicating gate functionality in experiment; the yellow outer barriers are all kept at constant voltage offset, red pixel gates are swept from \SI{-1.1}{\volt} to \SI{-1.5}{\volt} in 200 steps and define the QPC sweeping axis $V_\text{QPC}$, and blue pixel gates are optimized by the algorithm. (b) Loss for each iteration of the algorithm, dots are iterations where the best member of the population is shown in (d); the inset shows the optimized voltages for the best member of iteration 51 and 85. (c) Average voltage for each pixel per iteration, and variance across the population shown in shaded area. Colors correspond to $V_{\eta \nu}$ according to the inset. (d) Best-scoring member of each iteration, with later iterations appearing darker. The unoptimized result (blue) is the best evaluated staircase in the first iteration of the algorithm, the optimized staircase (light blue)  is the best ever seen result (iteration 51), and the dashed red staircase comes from iteration 85.}
    \label{fig:exp_results}
\end{figure*}


\paragraph*{Experimental setup}
\label{Experiments}
Finally, we turn to the experimental application of our technique on a device consisting of a pixelated QPC, similar to the simulated design, fabricated on a GaAs heterostructure (see Supplementary). The algorithmic capabilities demonstrated in the simulation were put to the test on a live device, with the goal of autonomously tuning multiple parameters to a desirable voltage region.
We use the same evolutionary optimization strategy as in the simulations and let the algorithm acquire measurements and control gate voltages without human intervention, but we optimized the exact tuning protocol to run smoothly on a real device.
In this section we outline the small changes made to the algorithm as well as the results obtained.

The device used is shown in Fig.~\ref{fig:exp_results}(a): \SI{18}{\nano\meter} thick gold gates were fabricated with a \SI{5}{\nano\meter} titanium adhesion layer on GaAs with a \SI{190}{\nano\meter} deep 2DEG (see Supplemetary). A set of outer gates was fabricated, designed to deplete the 2DEG and form a wide channel. In the middle of this channel, gold gates, consisting of pixels of size $300 \times 300\,$nm and pitch \SI{464}{\nano\meter} were fabricated, giving a total width of the QPC region of roughly \SI{1500}{\nano\meter}. Each pixel is connected through a \SI{30}{\nano\meter} thick gate-fanout, which is routed out to bond pads between the pixels and the outer gates. It should be noted that the need to make electrical connection to these pixels inherently breaks the symmetry of the device, and sets it apart from the idealized system we considered in our simulations, see Fig.~\ref{fig:exp_results}(a).

All measurements are carried out at \SI{100}{\milli\tesla} magnetic field applied out-of-plane, intended to aid QPC formation by creating edge states and reducing backscattering \cite{QPC_magnetic_field}. Conductance through the QPC region is measured with standard four-probe lock-in techniques in a dilution refrigerator at a few tens of millikelvins.

In order to fit all nine pixels and their electrical connections to the bond pads within fabrication constraints, the gate geometry of the fabricated device deviates from the simulated device in several aspects [see Fig.~\ref{fig:exp_results}(a)]. First, the two rectangular outer gates of the simulated device were replaced by eight tapered outer gates in the fabricated device, in order to make space for nine thin lines making connection between the pixel array and associated bond pads near the edge of the chip. Similar to the simulated devices, we fix the outer gates at values below their depletion voltage, and control the active QPC potential via the nine pixel-gate voltages. Second, spaces between pixel electrodes were made sufficiently large to accommodate the thin connecting lines where needed. This changes the spatial effects of each gate voltage and reduces device symmetry, possibly rendering the use of combined Fourier modes an unintuitive choice. Indeed, testing of the device using manual tuning via combined Fourier modes did not immediately give promising results, possibly aggravated by initial gate hysteresis.

Instead of using combined Fourier modes, we therefore apply $V_{\text{QPC}}$ to two opposing pixels (colored red in Fig.~\ref{fig:exp_results}a), i.e. $V_{\text{QPC}} = V_{-11} = V_{11}$, and let the algorithm freely optimize all other seven pixel voltages ($V_{-1-1}$, $V_{-10}$, $V_{0-1}$, $V_{00}$, $V_{01}$, $V_{1-1}$, $V_{10}$). The range allowed for $V_{\text{QPC}}$, and the choice of using the rightmost pixels $V_{-11}$ and $V_{11}$, was based on precharacterization using manual tuning, and we expect that other choices may also be suitable.

\paragraph*{Experimental results}
\label{Experiment_results}
The algorithm was allowed to autonomously optimize from a starting point of \SI{0}{\volt} for all pixel gates not used as $V_\text{QPC}$. 
Each pixel was bounded to $[\SI{-1.5}{\volt},\SI{0.3}{\volt}]$ to prevent device damage and hysteresis effects. 
The algorithm generally converged to regions with lower loss values than it started with. 
The loss of an example run is shown in Fig.~\ref{fig:exp_results}(b), where it drops gradually for the first 30 iterations, and then remains steady for the rest of this run. 
The inset shows examples of two of the best voltage configurations, runs 51 and 85 with losses of $36.14$ and $36.46$, respectively. 

We note the voltages keep drifting considerably after iteration 51 (the best solution) until the end of the optimization. 
This effect is even clearer from Fig.~\ref{fig:exp_results}(c) where pixel gate averages and variances for each iteration are plotted, showing that even towards the end of the run the algorithm is still exploring. 
There are several explanations for this behavior, including heating effects from measuring for extended periods during the optimization run and flat plateaus in the loss landscape where multiple configurations are equally good~\footnote{Ill-behaved gate hysteresis may also prevent convergence of the algorithm. To minimize gate hysteresis when measuring traces as in Fig.~\ref{fig:exp_results}(d), we hold the device at the initial gate voltages for at least $10$ seconds, before sweeping $V_{\text{QPC}}$.}.
A likely explanation is that it is due to measurement noise present in the system, which can turn the latest iterations of the optimization into an effective random walk; this occurs when the losses of individual members get so close to each other that the ranking used by the algorithm is in fact determined by measurement noise. 

In Fig.~\ref{fig:exp_results}(d) we show the results of the run in terms of the QPC traces obtained. 
The unoptimized trace (blue) is the lowest loss (best) member of the first iteration of the algorithm, showing no step-like conductance features. 
The final optimized version (light blue) is the best member of all iterations, here from iteration 51, where step-like features are clearly visible. 
We attribute this to the formation of a ballistic transport constriction, indicating a successful tuning of the QPC. 
A promising member (dashed red) from iteration 85 is also shown; interestingly it also appears with well defined conductance plateaus even though the pixel voltages are quite different from iteration 51 [see inset of panel (b) in Fig.~\ref{fig:exp_results}]. 
Each iteration has a population of nine members, meaning that the optimization found an optimum within 460 total measurements. 
 
The two configurations exemplified in Fig.~\ref{fig:exp_results}(d) (iterations $51$ and $85$) show excellent step behavior in their $G(V_{\text{QPC}})$ traces, yet their pixel configurations [insets of Fig.~\ref{fig:exp_results}(b)] indicate a significantly different landscape. 
This may indicate that the optimized voltages found by the algorithm are not concomitant to one QPC model potential, such as the canonical saddle-point potential.
This could be due to device asymmetry, and how the routing to the pixel gates influences the potential in the 2DEG. 
It could also mean that the algorithm optimizes pixel gate potentials to mitigate disorder potentials in the heterostructure or device, rather than defining a particular potential; this would be an important point for tuning of semiconductor quantum devices in general. 
For the particular case of QPC tuning, this could be alleviated by further measurements and improved device design, preferably low-impurity heterostructures.



\FloatBarrier

\paragraph*{Conclusions and outlook}
\label{Conclusion}
The results show that automated tuning of a QPC device with a $3\times 3$ array of tunable electrostatic gates using the CMA-ES algorithm gives more pronounced quantized conductance steps by optimizing the shape of the potential in the conductance channel. Through simulations we found that the algorithm is capable of adapting to random disorder on length scales comparable to the tunable gates. 
We furthermore demonstrated the CMA-ES algorithm on a real device, showing how it improved the conductance curves \textit{in situ}.
Although the actual experiment was limited to only tuning seven out of nine pixel gates, an improved device design featuring a multi-layer gate structure could possibly fix this issue in the future. Another outlook for future projects could be to allow voltage configurations that vary as a function of the sweeping parameter $V_{\text{QPC}}$, for instance by writing the Fourier coefficients $A_{\alpha\beta}=A_{\alpha\beta}(V_{\text{QPC}})$ as low-order polynomials where the prefactors can be tuned.


\paragraph*{Acknowledgements}
We acknowledge fruitful discussions with and contributions from A.~R.~Akhmerov. We also thank Xavier Waintal and Bernd Rosenow for initial discussions on this work and related projects.
This project was funded within the QuantERA II Programme that has received funding from the European Union’s Horizon 2020 research and innovation programme under Grant Agreement No 101017733.
It is part of INTFELLES-Project No.\ 333990, which is funded by the Research Council of Norway (RCN), and it received funding from the Dutch National Growth Fund (NGF) as part of the Quantum Delta NL programme.
Simulations were performed on resources provided by the NTNU IDUN/EPIC computing cluster~\cite{sjalander_epic_2022}. 
AC and TR acknowledge support from the Inge Lehmann Programme of the Independent Research Fund Denmark, and the US Army Research Office (ARO) under Award No. W911NF-24-2-0043. JB acknowledges support from the QuSpin Mobility Grant. CMM acknoledges a research grant (Project 43951) from VILLUM FONDEN. OK received funding via the Innovation Fund Denmark for the project
DIREC (9142-00001B).

The code for the optimization of simulated devices can be found at \href{https://github.com/jacobdben/qpcRL}{https://github.com/jacobdben/qpcRL}.


\bibliography{main_NEW.bbl}

\end{document}


\title{Supplementary Material:\\ Automated \textit{in situ} optimization and disorder mitigation\\ in a quantum device}

\author{Jacob Benestad}
\affiliation{Center for Quantum Spintronics, Department of Physics, Norwegian University of Science and Technology, NO-7491 Trondheim, Norway}
\author{Torbjørn Rasmussen}%
\affiliation{Center for Quantum Devices, Niels Bohr Institute, University of Copenhagen, 2100 Copenhagen, Denmark}
\affiliation{QuTech and Kavli Institute of Nanoscience, Delft University of Technology, 2600 GA Delft, The Netherlands}

\author{Bertram Brovang}
\affiliation{Center for Quantum Devices, Niels Bohr Institute, University of Copenhagen, 2100 Copenhagen, Denmark}
\affiliation{QuTech and Kavli Institute of Nanoscience, Delft University of Technology, 2600 GA Delft, The Netherlands}

\author{Oswin Krause}
\affiliation{Center for Quantum Devices, Niels Bohr Institute, University of Copenhagen, 2100 Copenhagen, Denmark}

\author{Saeed~Fallahi}
\affiliation{Department of Physics and Astronomy, Purdue University, West Lafayette, Indiana 47907, USA}
\affiliation{Birck Nanotechnology Center, Purdue University, West Lafayette, Indiana 47907, USA}

\author{Geoffrey~C.~Gardner}
\affiliation{Birck Nanotechnology Center, Purdue University, West Lafayette, Indiana 47907, USA}

\author{Michael~J.~Manfra}
\affiliation{Department of Physics and Astronomy, Purdue University, West Lafayette, Indiana 47907, USA}
\affiliation{Birck Nanotechnology Center, Purdue University, West Lafayette, Indiana 47907, USA}
\affiliation{Elmore Family School of Electrical and Computer Engineering, Purdue University, West Lafayette, Indiana 47907, USA}
\affiliation{School of Materials Engineering, Purdue University, West Lafayette, Indiana 47907, USA}
	
\author{Charles M. Marcus}
\affiliation{Center for Quantum Devices, Niels Bohr Institute, University of Copenhagen, 2100 Copenhagen, Denmark}

\author{Jeroen Danon}
\affiliation{Center for Quantum Spintronics, Department of Physics, Norwegian University of Science and Technology, NO-7491 Trondheim, Norway}

\author{Ferdinand~Kuemmeth}
\affiliation{Center for Quantum Devices, Niels Bohr Institute, University of Copenhagen, 2100 Copenhagen, Denmark}
\affiliation{Institute of Experimental and Applied Physics, University of Regensburg, 93040 Regensburg, Germany}

\author{Anasua Chatterjee}
\affiliation{Center for Quantum Devices, Niels Bohr Institute, University of Copenhagen, 2100 Copenhagen, Denmark}
\affiliation{QuTech and Kavli Institute of Nanoscience, Delft University of Technology, 2600 GA Delft, The Netherlands}

\author{Evert van Nieuwenburg}
\affiliation{Lorentz Institute and Leiden Institute of Advanced Computer Science, Leiden University, P.O. Box 9506, 2300 RA Leiden, The Netherlands}

\date{\today}

\maketitle

\section{KWANT simulation details}

We use a $500\times 300$ square lattice with a lattice constant $a = \SI{10}{\nano\meter}$.
The pixel gate dimensions are $\SI{280}{\nano\meter}\times \SI{280}{\nano\meter}$, with a gap between them of $\SI{70}{\nano\meter}$. The split gates are separated by $\SI{1.12}{\micro\meter}$, each extending $\SI{30}{\micro\meter}$ out of the scattering region, and each being $\SI{980}{\nano\meter}$ wide in the $x$-direction, see the sketch in Fig.~1(a) of the main text.

To avoid computationally intensive numerics, we ignore dielectric screening and consider only the bare electrostatic potential in the 2DEG, which for a rectangular patterned gate $i\in\{\text{SG1}, \text{SG2}, (\eta, \nu)\}$ located at $L_i < x < R_i$ and $B_i < y < T_i$ at a distance $d$ above the 2DEG takes the analytical expression \cite{davies_modeling_1995, foulk_theory_2024}
\begin{equation}
\begin{split}
    E_i(x,y) =  -V_i \Big[  {} & {} g(x-L_i, y-B_i) + g(x-L_i, T_i-y) \\
    {} & {} + g(R_i-x, y-B_i) + g(R_i-x, T_i-y)\Big],
\end{split}
\end{equation}
when applying a voltage $V_i$ to gate $i$, where
\begin{equation}
    g(x, y) = \frac{1}{2\pi}\arctan{\left(\frac{xy}{d\sqrt{x^2 + y^2 + d^2}}\right)}.
\end{equation}
The continuous functions $E_i(x,y)$ are evaluated at the position $(x,y)$ of each site in the $500 \times 300$ grid. Note that the effect of increasing $d$ is to smooth out the potential in the 2DEG; here we use $d=\SI{200}{\nano\meter}$.

\section{Simulated staircases for different disorder realizations}

Simulating five runs of the optimization without disorder results in two different configurations and their mirror images, as shown in \cref{fig:no_dis_collection}. Note that for all additional simulations in the supplement the CMA-ES time limit was set to $24$ hours, as the $48$ hour simulations in the main text indicated good convergence already at this point.

\begin{figure}
    \centering
    \includegraphics[width=0.9\linewidth]{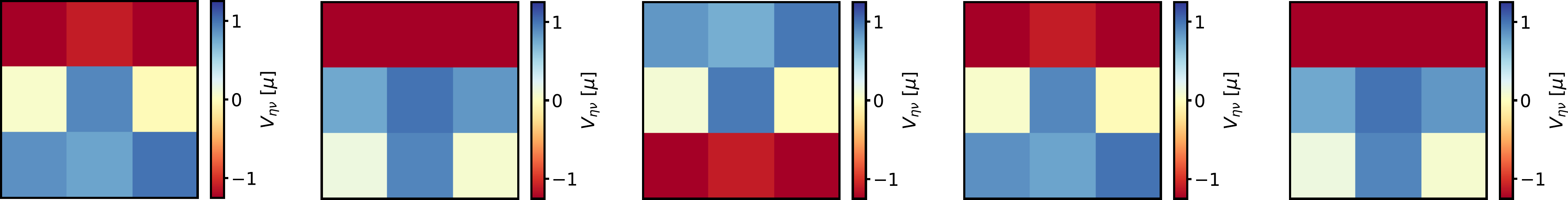}
    \caption{Resulting voltage configurations (subtracted by the average pixel value) of five simulations without the presence of disorder.}
    \label{fig:no_dis_collection}
\end{figure}

\begin{figure}
    \centering
    \includegraphics[width=\linewidth]{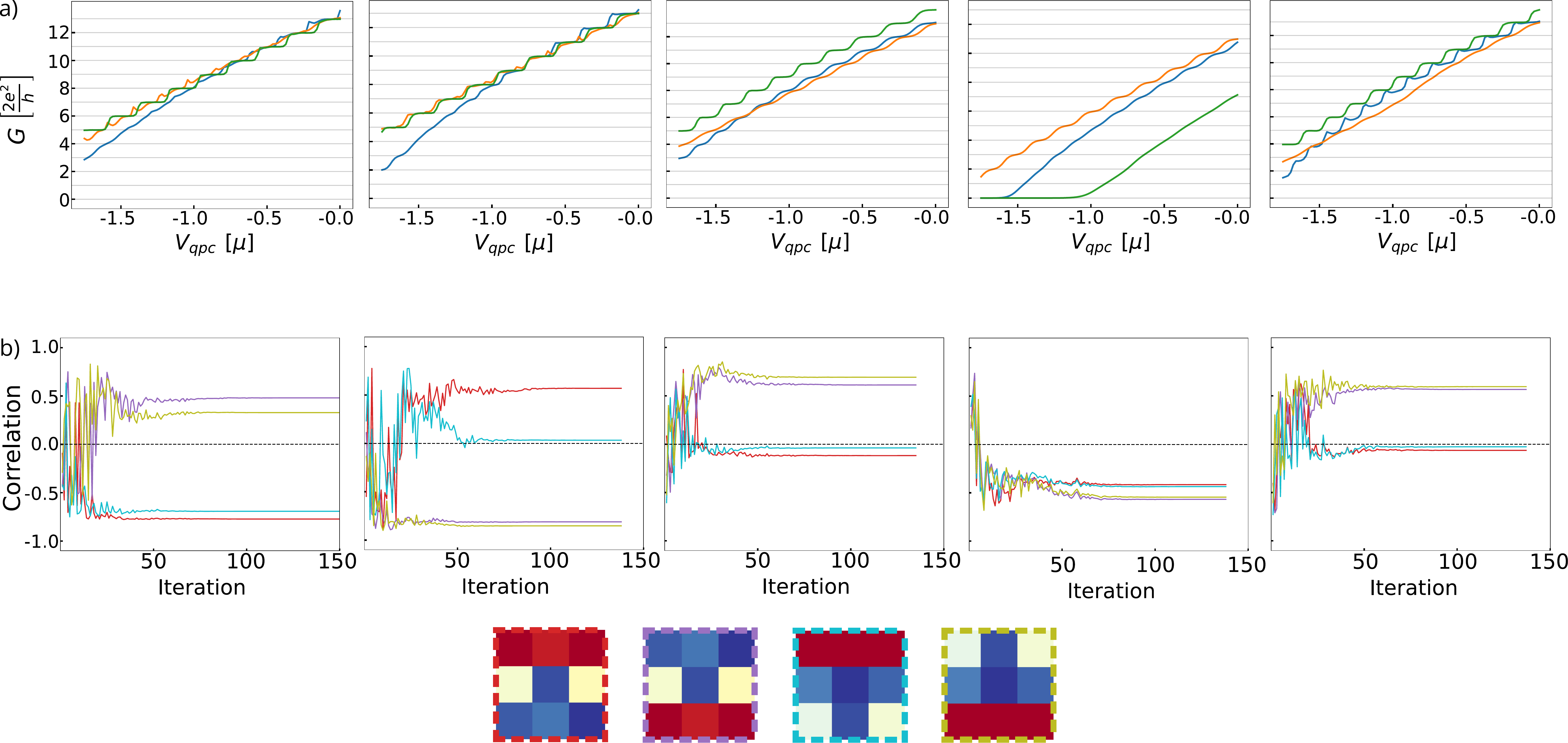}
    \caption{(a) Optimization results for different realizations of random disorder with correlation length $\lambda_{\text{dis}}=\SI{300}{\nano\meter}$ and magnitude $V_{\text{dis}}=0.352\,\mu$. Blue curves show the conductance when only sweeping the average gate value (no optimization), orange curves show sweeps with the solution found when optimizing without disorder, and green curves show the conductance when the average gate voltage is offset by the optimized solution for the disorder realization in question. (b) Correlation between the disorder potentials and optimized electrostatic potential differences in absence and presence of disorder. Each colored curve corresponds to the no-disorder solution depicted in the bottom four panels.}
    \label{fig:s1}
\end{figure}

To make sure that the simulation results presented in the main text also hold for other disorder realizations, we repeated the simulation five more times for different random configurations of disorder with the same correlation length. Resulting conductances for all these datasets with $\lambda_{\text{dis}}=\SI{300}{\nano\meter}$ are shown in \cref{fig:s1}(a), where the first panel presents the same data as used in the main text. Note that while the optimized solution in the presence of disorder mostly provides excellent improvement over solutions that do not consider the disorder, the optimization may on occasion fail as seen in the fourth panel of \cref{fig:s1}(a), where it in fact performs worse than the solutions that do not consider disorder.

In order to investigate if the optimized solution simply directly counteracts the disorder potential,
we also plot in \cref{fig:s1}(b) the average (across a conductance sweep) of the correlation between the actual disorder potential and ``correction potential'' resulting from the optimization,
\begin{equation}
    \text{Corr} \left[ E_{\text{dis}}(x,y), \, E_{\text{clean}}^{\text{opt}}(x,y)-E_{\text{dis}}^{\text{opt}}(x,y) \right] .
\end{equation}
Here $\text{Corr}\big[\cdot\big]$ is the real-space correlation in two dimensions, while $E_{\text{dis}}$, $E_{\text{clean}}^{\text{opt}}$ and $E_{\text{dis}}^{\text{opt}}$ are the disorder potential, optimized gate potential in the absence of disorder and optimized gate potential in the presence disorder, respectively. The four different curves correspond to the two solutions found when optimizing without disorder and their mirror images.
Note that in all cases except the one where the optimization fails, there is at least one instance of sizable positive correlation, which could mean that the optimized potential at least partly tries to compensate the disorder potential directly. However, the fact that there are four possible solutions to correlate with means one should be careful making definitive assumptions about this.

\begin{figure}
    \centering
    \includegraphics[width=\linewidth]{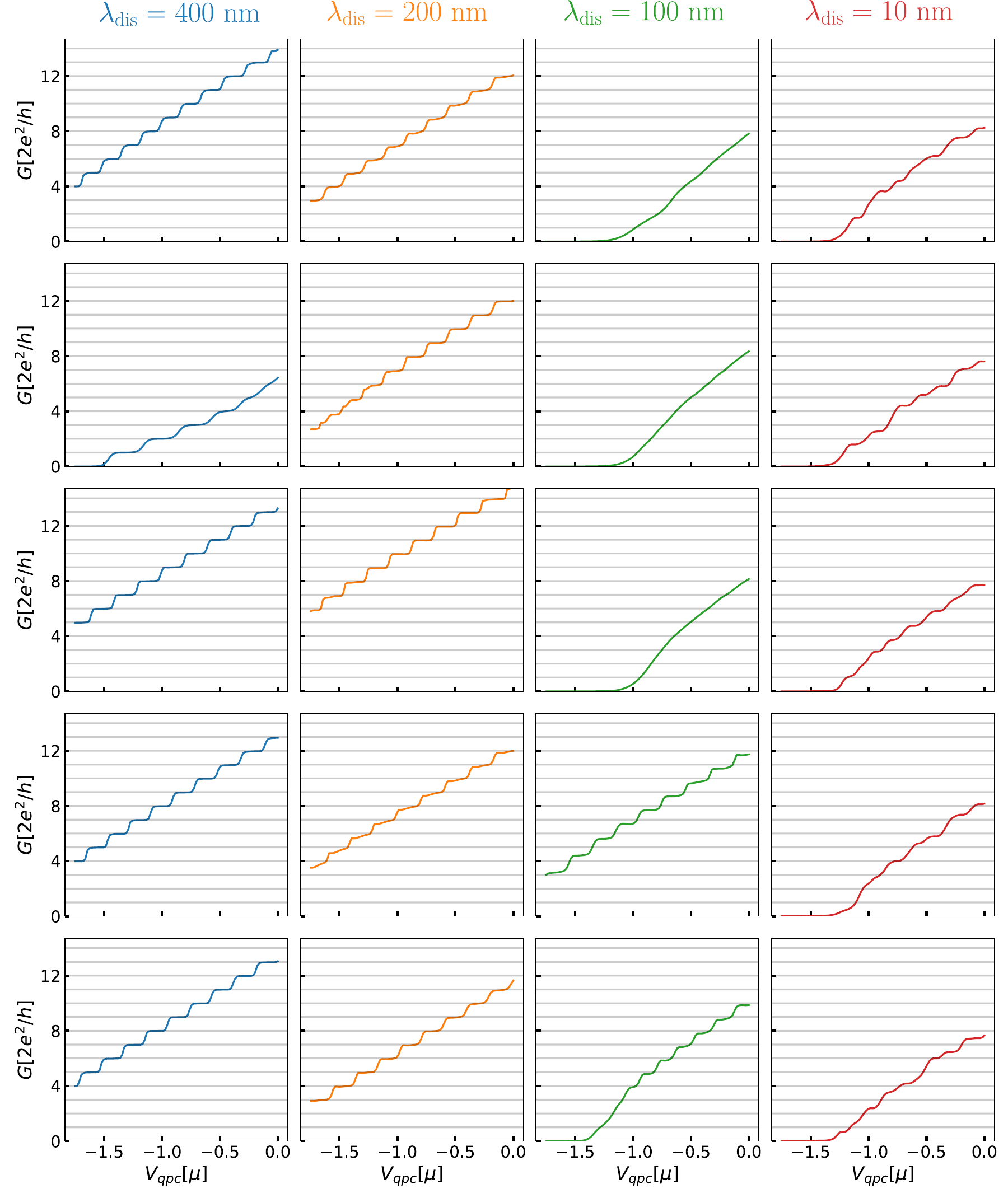}
    \caption{Conductances for optimized devices with different realizations of random disorder of magnitude $V_{\text{dis}}=0.352\,\mu$, with disorder length scales of $\lambda_{\text{dis}}=\SI{400}{\nano\meter}$, $\lambda_{\text{dis}}=\SI{200}{\nano\meter}$, $\lambda_{\text{dis}}=\SI{100}{\nano\meter}$ and $\lambda_{\text{dis}}=\SI{10}{\nano\meter}$ for the first, second, third and fourth column of panels respectively.}
    \label{fig:s2}
\end{figure}

\begin{figure}
    \centering
    \includegraphics[width=.5\linewidth]{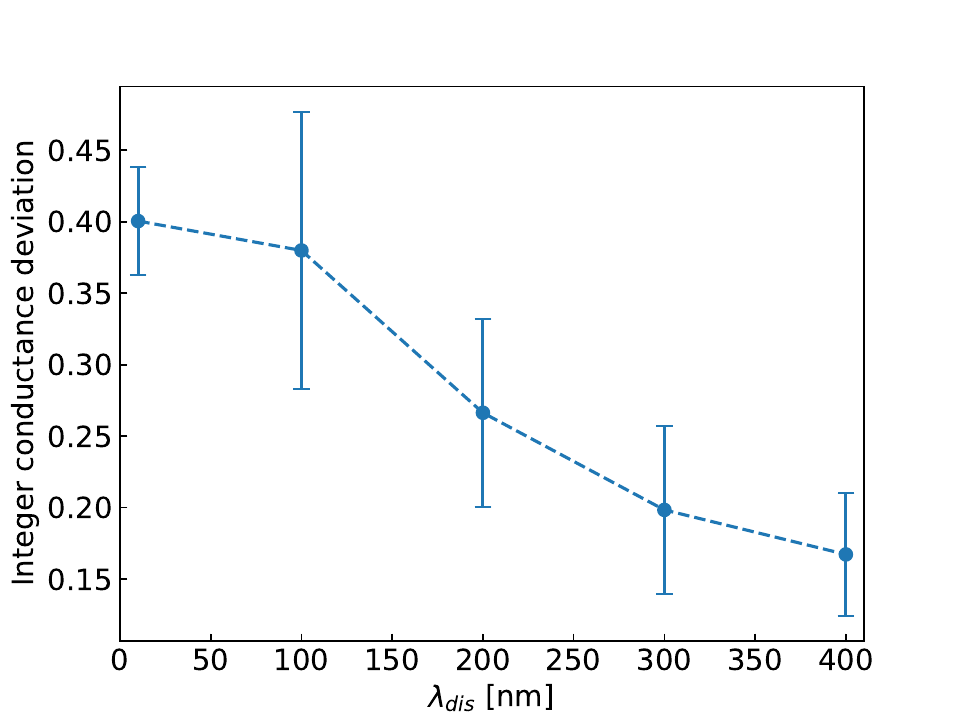}
    \caption{Ensemble average across five devices of two times the average difference from an integer value over a conductance sweep, with error bars extending one standard deviation to each side.}
    \label{fig:s3}
\end{figure}

Another point to investigate is how well the optimization performs for different disorder correlation lengths, as one would expect it to only work well for disorder sizes comparable to or larger than the pixels themselves. Repeating the simulations for disorder length scales of $\lambda_{\text{dis}}=\SI{400}{\nano\meter}$, $\lambda_{\text{dis}}=\SI{200}{\nano\meter}$, $\lambda_{\text{dis}}=\SI{100}{\nano\meter}$ and $\lambda_{\text{dis}}=\SI{10}{\nano\meter}$ (with again five different disorder realizations per correlation length) reveals that this does indeed seem to be the case. \Cref{fig:s2} shows the stepwise quantization deteriorate as the disorder length is reduced well below the pixel length. This deterioration is quantified in \cref{fig:s3}, where we plot $\text{E}_{\text{device}}\left\{\text{E}_{k}\left(2\big|G[k]- \big\lfloor G[k] \big\rceil \big|\right)\right\}$ with error bars extending one standard deviation to each side, i.e., $\pm \sqrt{\text{Var}_{\text{device}}\left\{\text{E}_{k}\left(2\big|G[k]- \big\lfloor G[k] \big\rceil \big|\right)\right\}}$, with the outermost expectation value and variance taken across the five devices with different disorder realizations, and the innermost expectation values across a conductance sweep (denoted here by indices $k$ corresponding to different values of $V_{\text{qpc}}$). 

\FloatBarrier
\section{Material and Device Fabrication}
The wafer used for the fabrication of the pixel QPC device is a GaAs/AlGaAs heterostructure. The 2DEG is \SI{190}{\nano\meter} below the surface and has a mobility of $24.6 \times 10^6~{\rm cm}^2/{\rm Vs}$. A mesa is first defined using electron beam lithography and then etched using a piranha etchant. Ohmic contacts are defined using electron beam lithography and the ohmic metal stack, consisting of platinum, gold and germanium, is deposited using electron beam evaporation. The ohmic metal stack is annealed into the 2DEG using a rapid thermal processing furnace. Finally the pixel QPC is defined using electron beam lithography and a metal stack of \SI{5}{\nano\meter} titanium and \SI{18}{\nano\meter} gold is deposited using electron beam evaporation.

\bibliography{supplement_NEW.bbl}